\documentclass[journal,twoside,web]{ieeeconf}

\usepackage{amsmath}
\usepackage{amssymb}
\usepackage{amsthm}

\usepackage{graphicx}
\usepackage{xcolor}

\usepackage{multirow}

\usepackage{caption}
\usepackage{subcaption}

\usepackage{hyperref}

\usepackage{tikz}
\usepackage{cite}
\usepackage{booktabs}   
\usepackage{xspace}

\usepackage{eso-pic}

\newtheorem{theorem}{Theorem}
\newtheorem{problem}{Problem}

\newtheorem{lemma}{Lemma}
\newtheorem{definition}{Definition}
\newtheorem{remark}{Remark}

\newcommand{\reals}{\mathbb{R}}

\newcommand{\expect}{\mathbb{E}}

\newcommand{\px}{\mathbf{x}}
\newcommand{\pu}{\mathbf{u}}
\newcommand{\pv}{\mathbf{v}}

\newcommand{\str}{\pi}

\newcommand{\U}{\mathcal{U}}

\newcommand{\G}{\mathcal{G}}
\newcommand{\F}{\mathcal{F}}

\newcommand{\I}{\mathcal{I}}
\newcommand{\D}{\mathcal{D}}
\newcommand{\N}{\mathcal{N}}

\newcommand{\V}{\mathcal{V}}
\newcommand{\X}{\mathcal{X}}

\newcommand{\fj}{f^{(j)}}
\newcommand{\faj}{\fj(\cdot, a)}

\newcommand{\DKL}{k_{dkl} (x_1, x_2) = k_{\gamma}(g^w_a(x_1), g^w_a(x_2))}
\newcommand{\SDDKL}{k_{dkl} (x_1, x_2) = k_{\gamma}(g^w_a(x_1)^{(j)}, g^w_a(x_2)^{(j)})}

\newcommand{\lowM}{\underline{M}_{q,a}}
\newcommand{\upM}{\overline{M}_{q,a}}
\newcommand{\lowS}{\underline{\Sigma}_{q,a}}
\newcommand{\upS}{\overline{\Sigma}_{q,a}}

\newcommand{\jth}[1]{#1^{(j)}}

\newcommand{\prop}{\mathrm{p}}
\newcommand{\gp}{{\texttt{GP}\xspace}}
\newcommand{\fd}{{\texttt{DKL}\!$^\texttt{F}$\xspace}}
\newcommand{\fdl}{{\texttt{DKL}\!$^\texttt{F\!L}$\xspace}}
\newcommand{\sd}{{\texttt{DKL}\!$^\texttt{S}$\xspace}}
\newcommand{\sdl}{{\texttt{DKL}\!$^\texttt{S\!L}$\xspace}}
\newcommand{\nngp}{{\texttt{NN-GP}\xspace}}
\newcommand{\nngpl}{{\texttt{NN-GP$^{\texttt{L}}$}\xspace}}

\newcommand{\err}{\text{err}}
\newcommand{\Dpred}{D_a^{\text{pred}}}
\newcommand{\map}{\mathbb{M}}

\DeclareMathOperator{\erf}{erf}
\DeclareMathOperator*{\argmin}{\arg \min}
\DeclareMathOperator*{\argmax}{\arg \max}
\DeclareMathOperator{\trace}{trace}

\title{\huge Promises of Deep Kernel Learning for Control Synthesis}

\author{Robert Reed$^{1}$, Luca Laurenti$^{2}$, and Morteza Lahijanian$^{1}$ 
\thanks{This work was supported in part by Air Force Research Lab (AFRL) under agreement number FA9453-22-2-0050.}
\thanks{$^{1}$Authors are with the Dept. of Aerospace Engineering Sciences at University of Colorado Boulder. {\tt\small \{Robert.Reed-1, Morteza.Lahijanian\}@colorado.edu }}%
\thanks{$^{2}$Author is with the Delft Center for Systems and Control, TU Delft {\tt\small L.Laurenti@tudelft.nl}}%
}

\begin{document}
\AddToShipoutPictureBG*{%
  \AtPageUpperLeft{%
    \hspace{15.2cm}%
    \raisebox{-1.5cm}{%
      \makebox[0pt][r]{Published in IEEE Control Systems Letters, December 2023.}}}}

\maketitle
\thispagestyle{empty}

\begin{abstract}
    Deep Kernel Learning (DKL) combines the representational power of neural networks with the uncertainty quantification of Gaussian Processes.  Hence, it is potentially a promising tool to learn and control complex dynamical systems. 
    In this work, we develop a scalable abstraction-based framework that enables the use of DKL for control synthesis of stochastic dynamical systems against complex specifications. Specifically, we consider temporal logic specifications and create an end-to-end framework that uses DKL to learn an unknown system from data and formally abstracts the DKL model into an interval Markov decision process to perform control synthesis with correctness guarantees. 
    Furthermore, we identify a deep architecture that enables accurate learning and efficient abstraction computation.
    The effectiveness of our approach is illustrated on various benchmarks, including a 5-D nonlinear stochastic system, showing how control synthesis with DKL can substantially outperform state-of-the-art competitive methods. 
\end{abstract}


\section{Introduction}
    \label{sec:intro}

Data-driven control synthesis is emerging as an important research topic in recent years~\cite{Dutta2018, Haesaert2017, Jackson2021, Nejati2023, abate2015}. 
This is due to three main reasons: (i) increased complexity of modern systems, which often include black-box components, (ii) availability of data in large scale, and (iii) increased capability of machine learning (ML) techniques.  There are however several challenges in data-driven approaches for control systems, especially in \emph{safety-critical} applications where robustness guarantees are vital. Such guarantees are conditioned on quantification of the learning error and its propagation through the control synthesis procedure.  While there exist ML techniques that supply information about the error \cite{Knuth2021}, they are often empirical (statistical) and lack necessary mathematical rigor.  Those methods that do provide formal error analysis \cite{Lederer2019} suffer from scalability~\cite{jackson2021formal, wajid2022formal}. This work focuses on these challenges and aims to provide a scalable data-driven control synthesis framework with robustness guarantees.

Formal synthesis is a rigorous approach to providing guarantees on the performance of control systems against complex properties~\cite{Tabuada2009, Lahijanian:TAC:2015}. In this approach, specifications are expressed in a formal language such as \textit{linear temporal logic} (LTL) over \textit{finite} behaviors (LTLf)~\cite{Giacomo2013} and the system progression is abstracted into a finite model called an \textit{abstraction}. 
Then, automated model-checking-like algorithms are used on the abstraction to synthesize a controller. 
To ensure correctness, the abstraction must have a \emph{simulation} relation with the system, which is often achieved by including all the uncertainties, e.g., errors due to discretization, stochasticity, and learning, in the abstraction.  A popular model that allows that is Interval Markov Decision Process (IMDP) \cite{givan2000bounded}, which is shown to also enable scalability to high dimensional systems~\cite{Cauchi:HSCC:2019}.  A key aspect in constructing a scalable IMDP abstraction is an accurate representation of the system evolution with tight uncertainty bounds.  That, however, is difficult to achieve in a data-driven setting.

A widely-used method for accurate representation of the latent control system from data is \emph{Gaussian Process} (GP) regression \cite{rasmussen:book:2006, jackson2021formal, Lederer2019}.  
Its power lies in rigorous uncertainty quantification, which comes at the expense of cubic computational complexity in the size of data. That makes GPs ideal for formal control synthesis, but they suffer in high dimensional spaces, where a massive amount of data is required to obtain small uncertainty.  
For high-dimensional systems, \emph{neural networks} (NNs) are successfully used to learn the dynamics, called \emph{NN dynamic models} (NNDMs)~\cite{liu2022},
with control synthesis methods~\cite{Adams:CSL:2022,Mazouz:Nurips:2022}. However, quantification of the learning error of NNDMs in a formal manner remains an open problem in spite of recent attempts to use confidence-based approaches~\cite{Knuth2021}, which cannot be propagated through the synthesis procedure.

In this work, we bridge the gap by introducing a scalable synthesis framework that harnesses the representational power of NNs and uncertainty quantification ability of GPs.  Specifically, we employ \emph{deep kernel learning} (DKL)~\cite{Wilson2015, Ober2021}, which uses NNs as informed priors for GPs while maintaining an analytical posterior, to efficiently construct (accurate) IMDP abstractions.
To ensure the correctness of the abstraction, 
we leverage recent techniques for linear relaxations of NNs~\cite{wang2021beta} 
and provide bounds on the mean and variance of the GP.  
Critically, we show that the optimization problems that bound the probabilities in the IMDP construction reduce to evaluations of a finite set of points on an analytical function, resulting in computational efficiency.  
Then, we employ existing tools~\cite{Lahijanian:TAC:2015}
to synthesize a strategy on the IMDP that maximizes the probability of satisfying a given LTLf specification and is robust against the learning error.  We prove that this strategy can be mapped to the underlying latent system with correctness guarantees.
We illustrate the efficacy of our framework on various benchmarks, which show control synthesis with DKL substantially outperforms state-of-the-art methods.  We also identify an architecture for DKL that results in high accuracy and efficiency in abstraction construction, promoting further scalability.

In summary, the contributions are: 
(i) a scalable data-driven framework for control synthesis with complex specifications and hard guarantees,
(ii) an efficient finite abstraction technique for DKL models with correctness guarantees,
(iii) a DKL architecture design for fast and accurate abstraction, and
(iv) illustration of the efficacy and scalability of the framework via benchmarking against state-of-the-art methods on a set of rich case studies with complex nonlinear systems up to 5 dimensions via deep architectures up to 3 hidden-layers and 100s of neurons.

\section{Problem Formulation}
    \label{sec:problem}
    Consider the following discrete-time stochastic system:
\begin{align}
    \label{true_dynamics} 
    \px(k + 1) &= f(\px(k), \pu(k))  + \pv(k), 
\end{align}
where $\px(k) \in \reals^n$, $\pu(k) \in U$, $U = \{a_1, \ldots, a_{|U|}\}$ is a finite set of actions or control laws, $\pv(k)\in \reals^n$ is a Gaussian random variable $\pv(k) \sim \N (0, \V)$ with zero mean and covariance $\V \in \reals^{n \times n}$, and $f: \reals^n \times U \rightarrow \reals^n$ is an \emph{unknown}, possibly non-linear, function. Without loss of generality, we assume covariance $\V$ is diagonal\footnote{There always exists a linear transformation, namely Mahalanobis transformation, that enables diagonalization of the covariance matrix.}. Intuitively, System~\eqref{true_dynamics} represents a switched stochastic systems with additive noise and unknown dynamics. 

We define a \textit{finite trajectory} of length $N \in \mathbb{N}$ of System~\eqref{true_dynamics} as  $\omega_{\px}^N = x_0 \xrightarrow{\pu_0} x_1 \xrightarrow{\pu_1} \cdots \xrightarrow{\pu_{N-1}} x_N$, where each $x_k \in \mathbb{R}^n$ is a sample from System~\eqref{true_dynamics}. 
We denote the $i$-th element of $\omega_{\px}^N$ by $\omega_{\px}^N \! (i)$ and
the set of all finite trajectories by $X^*$. A \textit{control strategy} $\str: X^* \rightarrow U$ is a function that chooses the next action $u \in U$ given a finite trajectory. Under $\str$ and initial condition $x_0 \in \reals^n$, System~\eqref{true_dynamics} defines a unique probability measure $P^{x_0}$ over $X^*$ \cite{bertsekas1996stochastic}.

We impose a standard smoothness (well-behaved) assumption on $f$.  
Namely, we assume $f$ is a sample from a Gaussian process (GP)\footnote{Note that the restrictions that this assumption poses on $f$ depends on the choice of the covariance (kernel) function for the GP, and there exist universal kernels, such as the squared exponential, that allow for a GP to approximate \emph{any} continuous $f$ arbitrarily well.} (see Sec. \ref{sec;GPPrelim} for details).
Since $f$ is unknown, we aim to reason about System~\eqref{true_dynamics} solely from a set of input-output data.  
Specifically, we assume $D = \{(x_i,u_i,x_i^+)\}_{i=0}^m$ is a set of identically and independently distributed (i.i.d.) data, where ${x}_i^+$ is a sample of a one time-step evolution of System~\eqref{true_dynamics} from $x_i \in \reals^n$ under action $u_i \in U$.

We are interested in the temporal properties of $\px$ in a compact set $X \subset \reals^n$ w.r.t. a set of regions $R = \{r_1, \ldots, r_l\}$, where $r_i \subseteq X$. To this end, we define a set of atomic proposition $\Pi = \{\prop_1, \ldots, \prop_l\}$, where $\prop_i$ is true iff $\px \in r_i$.  Let $L: X \to 2^\Pi$ be a labeling function that assigns to each state the set of atomic propositions that are true at that state.  Then, the observation trace of trajectory $\omega_\px^N$ is $\rho = \rho_0 \rho_1 \ldots \rho_N$, where $\rho_i = L(\omega_\px^N \! (i))$ for all $0 \leq i \leq N$.  

To express the temporal properties of System~\eqref{true_dynamics}, we use LTLf~\cite{Giacomo2013}, which has the same syntax as LTL but its interpretations are over finite behaviors (traces).

\begin{definition}[LTLf]
    \label{def:LTLf}
    Given a set of atomic propositions $\Pi$, an LTLf formula is defined recursively as
    \begin{equation*}
        \varphi = \prop \mid \neg \varphi \mid \varphi \land \varphi \mid \bigcirc \varphi \mid  \varphi \U \varphi \mid \F \varphi \mid \G \varphi
    \end{equation*}
    where $\prop \in \Pi$, and $\bigcirc$, $U$, $\F$, and $\G$ are the ``next'', ``until'',  ``eventually'', and ``globally'' temporal operators, respectively.
\end{definition}
The semantics of LTLf are defined over finite traces \cite{Giacomo2013}. We say trajectory $\omega_\px \in X^*$ satisfies formula $\varphi$, denoted by $\omega_\px \models \varphi$, if a prefix of its observation trace satisfies $\varphi$.

\begin{problem} [Control Synthesis] \label{prblm:1} 
    Given a dataset $D = \{({x}_i, {u}_i, {x}_i^+)\}_{i=1}^{m}$ of i.i.d. samples of System~\eqref{true_dynamics}, compact set $X$, and LTLf formula $\varphi$, find control strategy $\str^*$ that maximizes the probability of satisfying $\varphi$ without existing $X$, i.e., for every $x_0 \in X$, 
    \begin{align}
        \pi^* = \arg\max_\pi  P^{x_0}(\omega_\px \models \varphi \mid D, \str)
    \end{align}
\end{problem}

There are three main challenges in Problem \ref{prblm:1}: (i) the dynamics of System~\eqref{true_dynamics} are unknown, can be nonlinear, and its evolution is stochastic, (ii) guarantees are required for the underlying system to satisfy complex specifications, and (iii)
scalability to higher dimensions is necessary, which is an additional challenge that we impose. 
In our approach, we show that challenges (i) and (iii) can be successfully addressed by utilizing the power of DKL to approximate $f$ at the low level. For challenges (ii) and (iii), we draw inspirations from formal methods literature and construct a discrete abstraction of the dynamics as an IMDP. With an IMDP and an LTLf specification, we can use off-the-shelf tools for synthesizing provably correct strategies.

\section{Modelling Dynamical Systems using Deep Kernel Learning}
    \label{sec:modeling_DKL}
    
To describe how we employ DKL to learn $f$ in System~\eqref{true_dynamics}, we first need to introduce GPs. Then, we present DKL within the GP framework.

\subsection{Gaussian Process Models}
\label{sec;GPPrelim}
A Gaussian process (GP) is a collection of random variables, such that any finite collection of those random variables are jointly Gaussian \cite{rasmussen:book:2006}. Because of the favorable analytical properties of Gaussian distributions, GPs are widely employed to learn unknown functions, such as $f$ in System~\eqref{true_dynamics}, from observations of the system~\cite{pmlr-v37-sui15, Lederer2019}. In particular, given a prior GP, $\textrm{\textit{GP}}(\mu, k_{\gamma})$, where $\mu:\mathbb{R}^{n}\to\mathbb{R}$ is the mean function and $k_{\gamma}:\reals^{n} \times \reals^{n} \to \reals$ is a positive semi-definite covariance function (or kernel) with hyper-parameters $\gamma$, the assumption is that for each $a\in U$ and for each $j\in\{1,\ldots,n\}$, $\faj$, the $j$-th component of $f(\cdot, a)$, is a sample from $\textrm{\textit{GP}}(\mu, k_{\gamma})$.
Then, given dataset  $D = \{(x_i,u_i,x_i^+)\}_{i=0}^m$ of samples of System~\eqref{true_dynamics}, which we partition in $|U|$ subsets
$D_a = \{({x}, u, {x}^+) \in D \mid u = a\}$, we obtain that, at every point $x^* \in \reals^n$, the posterior predictive distribution of $f^{(j)}(x^*,a)$ given $D$ is still Gaussian with mean and variance: 
\begin{align}
    &\expect(\fj(x^*, a)\mid D) = \nonumber \\ 
    & \hspace{25mm}
    \mu(x^*) + K_{x^*, \X}(K_{\X,\X} + \sigma^2 I)^{-1}Y, \label{mean_pred}\\
\nonumber
    &cov(\fj(x^*, a)\mid D) = \\
    & \hspace{18mm} K_{x^*, x^*} - K_{x^*,\X}(K_{\X,\X} + \sigma^2 I)^{-1}K_{\X, x^*}, \label{covar_pred}
\end{align}
where $\X = (x_1, \ldots, x_{|D_a|})$, $Y = (x_1^{(j)+}, \ldots, x_{|D_a|}^{(j)+})$, and $K_{\X,\X} \in \reals^{|D_a|\times |D_a|}$
is a matrix whose $i$-th row and $l$-th column is $ k_{\gamma}(x_i, x_l)$.

A widely-used kernel function is the squared exponential:
\begin{align}
    k_{\gamma_{se}}(x,x') = \sigma_{s} \exp \left(\frac{-\|x-x'\|}{2 l^2}\right)
\end{align}
with the set of hyper-parameters $\gamma_{se} = \{\sigma_s, l\}$, where $\sigma_s$ and $l$ are the output scale and length scale, respectively. These hyper-parameters are generally learned by minimizing the negative marginal log-likelihood of the data \cite{rasmussen:book:2006}. 

\subsection{Deep Kernel Learning} \label{subsec:DKM}

\begin{figure}[t]
        \centering
    \begin{subfigure}[b]{0.45\linewidth}
        \includegraphics[width=\linewidth]{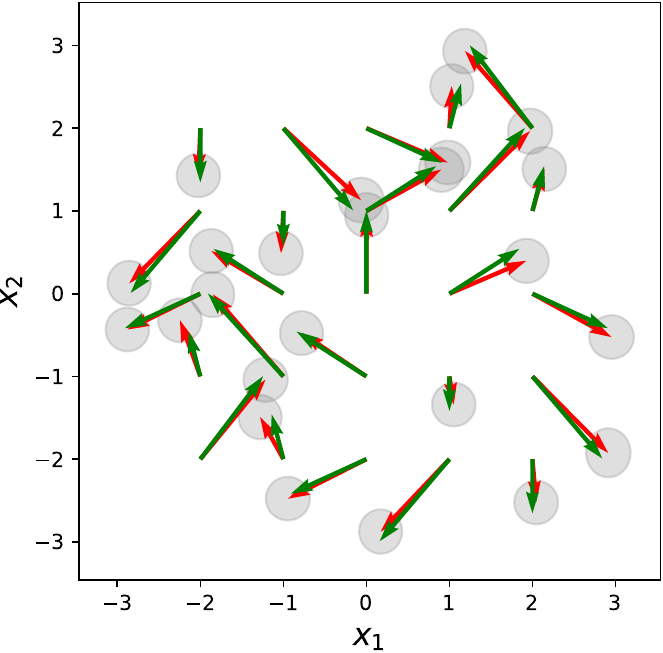}
    \end{subfigure}
    \hfill
    \begin{subfigure}[b]{0.45\linewidth}
        \includegraphics[width=\linewidth]{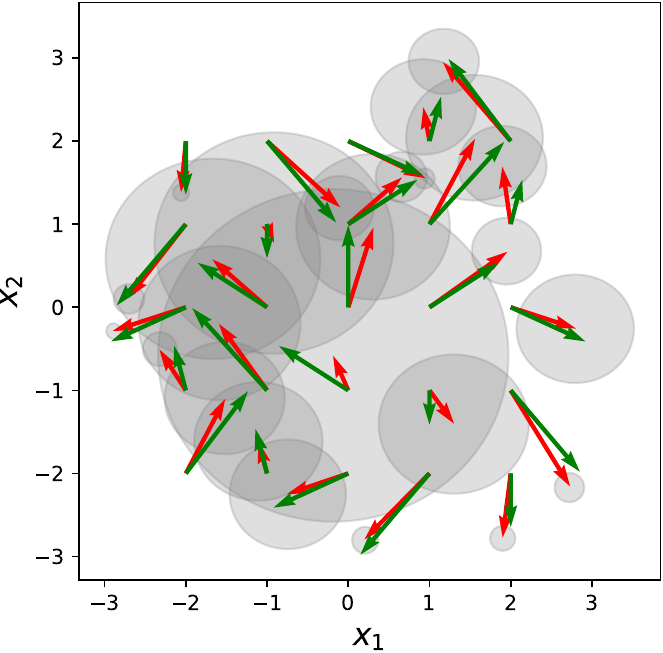}
    \end{subfigure}
    \begin{subfigure}[b]{0.45\linewidth}
        \includegraphics[width=\linewidth]{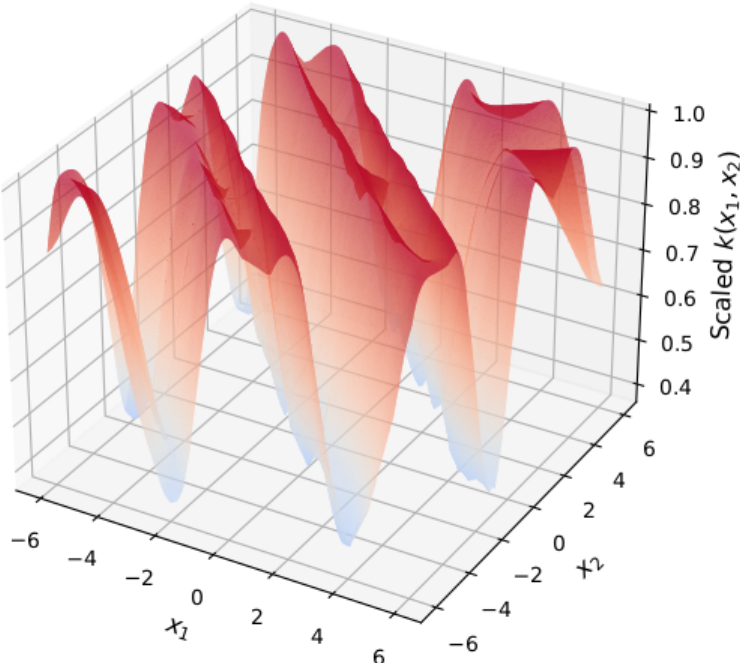}
    \end{subfigure}
    \hfill
    \begin{subfigure}[b]{0.07\linewidth}
        \includegraphics[width=\linewidth]{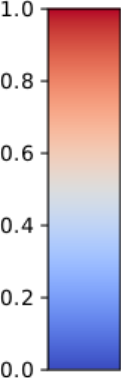}
        \vspace{3mm}
    \end{subfigure}
    \hfill
    \begin{subfigure}[b]{0.45\linewidth}
        \includegraphics[width=\linewidth]{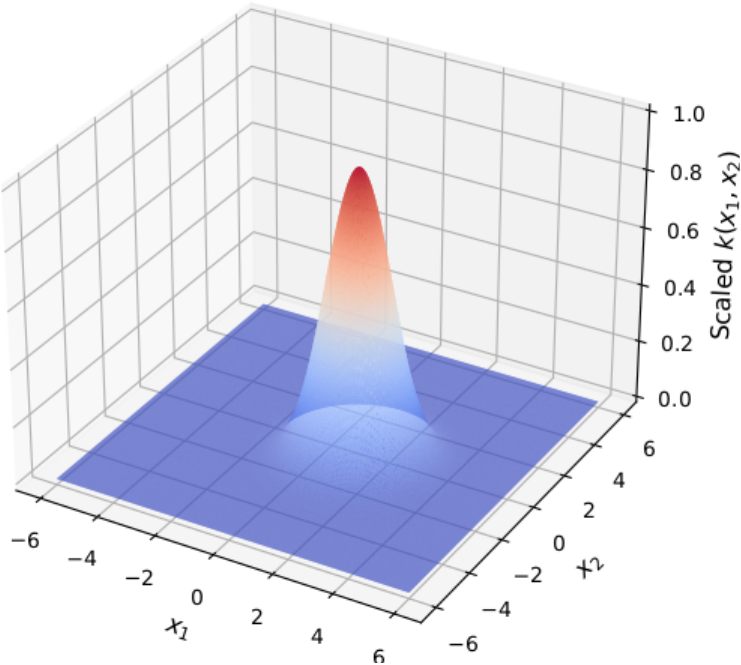}
    \end{subfigure}
    \caption{Results on learning a 2D vector field with (left) DKL and (right) GP. 
    Top: the true vector field in green, the model predictive posteriors in red, and 95\% confidence intervals shaded in grey. Bottom: scaled correlation function for the first dimension at one point. 
    1000 samples were used to pre-train the NN. Both methods use 100 samples for predictions.}
    \label{fig:dkl_vs_gp}
\end{figure}

The squared exponential kernel, like most commonly-employed kernels for GP regression~\cite{rasmussen:book:2006}, only depends on few hyper-parameters. This limits the flexibility of GPs in learning complex representations of data \cite{calandra2016manifold}, often resulting in predictions with large uncertainty (variance). One can reduce this uncertainty with more data, but that leads to computational intractability since the time complexity of GP regression is $\mathcal{O}(|D|^3)$~\cite{rasmussen:book:2006}. 
DKL aims to address this issue by considering a kernel that is composed with a NN. The underlying idea is that a fully connected NN $g^w_a:\mathbb{R}^n \to \mathbb{R}^s$, parameterized by weights and biases vector $w$ over action $a$, is employed to map the input into an $s$-dimensional feature space, where GP regression is performed. 
Specifically, starting with a base kernel $k_{\gamma}$ (in this paper we always assume $k_{\gamma}=k_{\gamma_{se}}$), we define a deep kernel as 
\begin{align}
    \label{eq:deep kernel}
    k_{dkl}(x,x') = k_{\gamma}(g^w_a(x), g^w_a(x')).
\end{align} 
Then, with $k_{dkl}$, predictions still use GP's mean and covariance equations in \eqref{mean_pred}-\eqref{covar_pred}, but the number of hyper-parameters (i.e., $\gamma$ and $w$) are drastically increased.
This significantly improves flexibility and representational power of GPs.

The learning of the parameters in $\gamma$ and $w$ can be achieved by either minimizing the negative marginal log-likelihood or considering a fully Bayesian approach~\cite{Ober2021}. 
Furthermore, the NN portion of DKL models can be pre-trained and its parameters fixed. This minimizes the number of parameters being optimized through the marginal log-likelihood and mitigates the possibility of DKL over-fitting the data~\cite{Ober2021}.

DKL combines the flexibility of deep NN with the principled uncertainty quantification of GPs. Such a combination is particularly important for problems that require learning complex non-linear dynamics with robustness analysis.  The former often necessitates large amount of data and the latter reasoning about uncertainty.
The power of DKL is illustrated in Figure~\ref{fig:dkl_vs_gp}, where we consider learning a 2D vector field $f(\px) = (\sin{(x_1 + x_2)}, \cos{(x_1 - x_2)})^T$ with noise distribution $\mathcal{N}(0, 0.01)$ by using both a standard GP and DKL. We can observe that the $k_{dkl}$ is able to learn the oscillatory behavior of the data, while the GP with $k_{\gamma_{se}}$ only correlates nearby points. As a consequence, with the same amount of data, the predictions of DKL are more accurate and less uncertain compared to the ones of the GP. 
Particularly, in our synthesis framework, which relies on abstraction-based techniques, the lower uncertainty associated with DKL can lead to less conservative abstractions and probabilistic guarantees.

\begin{remark}
    We note that care must be taken for both architecture design and training of the NN portions of DKL.
    If the prior is trained poorly, the kernel may underestimate uncertainty, resulting in an inaccurate model.
    Also, architecture of the NN can have a significant role for both computational tractability and accuracy of the abstraction.
    In Sec.~\ref{sec:conclusion}, we compare a few architectures and discuss an appropriate form for control synthesis.  As for training technique, our empirical results show that stochastic mini-batching \cite{Ober2021} is highly effective. 
\end{remark}

\section{IMDP Abstraction}
    \label{sec:method}

DKL allows one to predict the one-step evolution of System~\eqref{true_dynamics} from a given $x\in X$ and $u \in \U$.  
To analyze LTLf properties of System~\eqref{true_dynamics}, however, we need to reason over finite trajectories (with arbitrary lengths) of System~\eqref{true_dynamics} and consequently perform multi-step predictions of arbitrary length. Unfortunately, such analysis is already intractable for standard GPs even for a fixed finite horizon \cite{deisenroth2011pilco}. To address this problem we rely on finite abstractions, which in turn allows one to use  existing LTLf control synthesis tools~\cite{Lahijanian:TAC:2015,Laurenti:TAC:2020}.  Specifically, we use an Interval Markov Decision Process (IMDP)~\cite{givan2000bounded} as the abstraction model due to its ability to represent multiple levels of uncertainty. 
 
\begin{definition} [IMDP] \label{def:imdp}
    An interval Markov Decision Process (IMDP) is a tuple $\I = (Q,A, \Sigma, \hat{P}, \check{P}, \Pi, L)$, where 
    \begin{itemize}
        \item $Q$ is a finite set of states,
        \item $A$ is a finite set of actions,
        \item $\hat{P}: Q \times A \times Q \rightarrow [0, 1]$ is a transition probability function that defines the upper bound of the transition probability from state $q\in Q$ to state $q'\in Q$ under action $a\in A$,
        \item $\check{P}: Q \times A \times Q \rightarrow [0, 1]$ is a transition probability function that defines the lower bound of the transition probability from state $q\in Q$ to state $q'\in Q$ under action $a\in A$,
        \item $\Pi$ is a set of atomic propositions, and
        \item $L: Q \to 2^\Pi$ is a labeling function that assigns to each state $q \in Q$ a subset of $\Pi$.
    \end{itemize}
\end{definition}
\noindent
It holds for all $q,q' \in Q$ and $a \in A(q)$ that $\check{P}(q,a,q') \leq \hat{P}(q,a,q')$ and 
$\sum_{q'\in Q} \check{P}(q,a,q') \leq 1 \leq \sum_{q'\in Q} \hat{P}(q,a,q').$  A \textit{finite path} of $\I$, denoted by $\omega_\I \in Q^*$, is a sequence of states in $Q$. A \textit{strategy} of $\I$ is a function $\str_\I: Q^* \to A$ that maps $\omega_\I$ to an action in $A$. 

\subsection{Building the Abstraction}

\subsubsection{States and Actions}

First, we partition $X$ into a set of convex regions $\bar{Q} = \{q_1, \ldots, q_{|\bar{Q}|}\}$, e.g., by using a grid. 
We consider an additional region $q_u = \reals^n\setminus X$ and call $Q=\bar{Q}\cup \{q_u\}$ the set of IMDP states.
We assume that the discretization $\bar{Q}$ of $X$ respects the regions of interest in $R$, i.e., $\forall r \in R$, $\exists Q_r \subseteq \bar{Q}$ such that $\cup_{q \in Q_r} q= r$. 
With an abuse of notation, we use $q$ to denote both a state in the IMDP and its corresponding region, i.e., $q \in Q$ and $q \subset \mathbb{R}^n$. Note that for every $x,x'\in q$, $L(x) = L(x')$; accordingly, we set the IMDP labeling function as $L(q) = L(x)$. 
The set of IMDP actions $A$ is given by the set of actions $U$ and all actions are allowed to be available at each state $q \in Q$.

\subsubsection{Probability Bounds}
The key step to building an IMDP abstraction of System~\eqref{true_dynamics} is the computation of the transition probability functions $\hat{P}$ and $\check{P}$. 
Given $q \subset \mathbb{R}^n$, $a \in U$, and $x \in X$, we define the \textit{transition kernel} $T_a(q \mid x)$ as:
\begin{multline}
    \label{Eqn:Kernel}
    T_a(q \mid x) = \int_{q}\N(v\mid \expect(f(x, a) \mid D), \\
    cov(f(x, a) \mid D) + \V)dv
\end{multline}
That is, $T_a(q \mid x)$ is  the probability that, given the data $D$, our Gaussian prior assumption on $f$, and an initial state $x$,  System~\eqref{true_dynamics} transitions to $q$ under $a$ in one time step. Note that $T_a(q \mid x)$  is defined by marginalizing the DKL predictive distribution for $f$ over the dynamics of System~\eqref{true_dynamics} and the resulting kernel is still Gaussian due to the closure of Gaussian random variables under linear combinations~\cite{rasmussen:book:2006}. As we show in Theorem \ref{thm:Correctness} in Sec.\ref{sec:ver-synth}, this marginalization guarantees that our abstraction accounts for the uncertainty coming from the DKL predictions. 

Consequently, for $q,q' \in \bar{Q}$, it follows that
\begin{align}
\label{eqn:lowerBuund}
     \check{P}(q, a, q') &= \min_{x \in q}  T_a(q' \mid x),\\
     \label{eqn:upperBound}
     \hat{P}(q,a,q') &= \max_{x \in q}  T_a(q' \mid x),
\end{align}
and for the unsafe region $q_u$, it holds that 
\begin{align}
     \check{P}(q, a, q_u) &= 1 - \max_{x \in q}  {T}_a(X \mid x), \\
     \hat{P}(q,a, q_u) &= 1 - \min_{x \in q}  {T}_a(X \mid x).
\end{align}
Lastly, since reaching $q_u$ violates the requirement of not leaving $X$, we set $q_u$ to be a sink state, i.e., $\forall a\in A$, $\check{P}(q_u,a,q_u)=\hat{P}(q_u,a,q_u)=1$. 

In the remainder of this section, we show how to efficiently compute the bounds in~\eqref{eqn:lowerBuund}-\eqref{eqn:upperBound}.
We start by noticing that local linear relaxations for NNs can be built in constant time by utilizing algorithms in~\cite{wang2021beta}. 
That is,  for NN $g^w_a$ and region $q \subset \reals^n$, one can find matrices $\check{A}_q,\hat{A}_q \in \reals^{n \times n}$ and $\check{b}_q,\hat{b}_q \in \reals^n$ such that $\forall x\in q$ it holds that:
\begin{align} 
    \label{eq:linear_relax}
    \check{A}_q x + \check{b}_q \leq g^w_a(x) \leq \hat{A}_q x + \hat{b}_q.
\end{align}
We use such relaxations to propagate $q$ through the NN of a deep kernel to produce an axis-aligned hyper-rectangle $Z_{q,a}$ that contains the output of $g^w_a(x)$ for every $x\in q.$ Then, given $Z_{q,a}$, we can use existing results for GPs~\cite[Propositions 4 and 7]{Patane2022} to propagate $Z_{q,a}$ through the squared exponential function and obtain the ranges of posterior mean and variance for all $x\in q$ by solving convex optimization problems, namely one quadratic program and three linear programs.
Then, we obtain mean bounds $\lowM,\upM \in \reals^n$  and variance bounds $\lowS,\upS \in \reals^{n}_{\geq 0}$ such that, for every $x \in q$ and every $j \in \{1,\ldots, n\}$, 
\begin{align}
\label{eqn:RangesMean}
    \expect(\fj(x, a)\mid D_a) \in [\jth{\lowM},  \jth{\upM}], \\
    cov(\fj(x, a)\mid D_a) \in [\jth{\lowS}, \jth{\upS}].
    \label{eqn:RangesVariance}
\end{align}



Using these bounds, in the following theorem, we show how to efficiently compute bounds for \eqref{eqn:lowerBuund}-\eqref{eqn:upperBound}

\begin{theorem}[Efficient Computation for Tran. Prob. Bounds]
    \label{th:GPBOunds}
    For $\mu \in \reals$ and $\sigma \in \reals_{\geq 0}$ and closed interval $\theta= [\underline{\theta}, \overline{\theta}] \subset \reals$,
    define function
    \begin{align*}
        h(\theta, \mu, \sigma) = \frac{1}{2} \left( \erf \left( \frac{\overline{\theta} - \mu}{\sqrt{2 \sigma}} \right) - \erf \left( \frac{\underline{\theta} - \mu}{\sqrt{2 \sigma}} \right) \right).
    \end{align*}
    Further, given region $q \in \bar{Q}$, let $[\lowM, \upM]$ and $[\lowS, \upS]$ be the poster mean and variance of DKL as reported in \eqref{eqn:RangesMean}-\eqref{eqn:RangesVariance}. 
    Additionally, for region $q' \in \bar{Q}$, denote its centroid  by $c_{q'}$ and define points
    \begin{align*}
        \underline{z} = \argmin_{z\in [\lowM, \upM]} \|z - c_{q'}\|, \\
        \overline{z} = \argmax_{z\in [\lowM, \upM]} \|z - c_{q'}\|.
    \end{align*}
    Then,  denoting the closed interval obtained by projecting $q' \subset \reals^n$ onto the $j$-th dimension by $\jth{q'} \subset \reals$, it holds that
    \begin{align*}
        \min_{x \in q}  T_a(q' \mid x) \geq \prod_{j=1}^n \min_{\tau \in \{\jth{\lowS}, \jth{\upS}\}} \!\!\! h(\jth{q'}, \jth{\overline{z}}, \tau + \V^{(j,j)}), \\
        \max_{x \in q}  T_a(q' \mid x) \leq \prod_{j=1}^n \max_{\tau \in \{\jth{\lowS}, \jth{\upS}\}} \!\!\! h(\jth{q'}, \jth{\underline{z}}, \tau + \V^{(j,j)}), 
    \end{align*}
    where $\V^{(j,j)}$ is the $j,j$ element of the noise covariance matrix $\V$.
\end{theorem}
\begin{proof}
In the proof, we consider the $\min$ case; the $\max$ case follows similarly.
Note that $h(\theta,\mu,\sigma)$ is the integral of $\N(\mu,\sigma)$ over $\theta$. Then,
under the assumption of diagonal $ cov(\fj(x, a)\mid D_a)$ and $\mathcal{V}$, it holds that for every $x\in q$,
\begin{align*}
&    T_a(q' \mid x)=  \prod_{j=1}^n  h(\jth{q'}, \expect(\jth{f}(x, a) \mid D), \\
& \qquad \qquad  \qquad \qquad  cov(\jth{f}(x, a) \mid D) + \V^{(j,j)}) \geq \\
 & \quad   \prod_{j=1}^n \min_{\mu \in [\lowM^{(j)}, \upM^{(j)}], \, \tau \in [\jth{\lowS}, \jth{\upS}]} \!\!\! h(\jth{q'}, \mu, \tau + \V^{(j,j)}).
\end{align*} 
Consequently, what is left to show is how to place mean $\mu$ and variance $\tau$ of a uni-dimensional Gaussian to minimize its integral over the respective dimension of $q'$. Each of these is minimized by first maximizing the distance of $\jth{z}$ from $\jth{c_{q'}}$, hence $z$ can be chosen according to  $\argmax_{z\in [\lowM, \upM]} \|z - c_{q'}\|$. Then, there are two cases: $\jth{z} \in \jth{q'}$ and $\jth{z} \not\in \jth{q'}$. In the first case, $T$ is minimized if we minimize the probability mass in $q'$, which results in $\tau=\jth{\upS}$. In the second case, with a similar reasoning we obtain $\tau=\jth{\lowS}$ or $\tau=\jth{\upS}$. 
\end{proof}

Theorem~\ref{th:GPBOunds} shows that 
we can compute transition bounds $\check{P}$ and $\hat{P}$ by simply evaluating an error function at $4 n$ points, thus guaranteeing efficient abstraction construction.

\section{Control Synthesis and Refinement}
    \label{sec:ver-synth}
    
Once we obtain IMDP abstraction $\I$, our goal is to synthesize a strategy $\str_\I$ that maximizes the probability of satisfying specification $\varphi$ on $\I$ and then map it back to System~\eqref{true_dynamics} to obtain control strategy $\str$.  

Let $\D(Q)$ be the set of all probability distributions over $Q$.
We define an \emph{adversary} $\nu_\I: Q^* \times A \to \D(Q)$ to be a function that maps a finite path $\omega_{\I} \in Q^*$ and an action $a \in A$ to a transition probability distribution such that, $\forall q' \in Q$, 
$$\check{P}(\text{last}(\omega_{\I}),a,q') \leq \nu_\I(\omega_{\I},a)(q') \leq \hat{P}(\text{last}(\omega_{\I}),a,q'),$$
where $\text{last}(\omega_{\I})$ is the last state in $\omega_\I$.
Given $\str_\I$ and $\nu_\I$, a probability measure $Pr$ over paths in $Q^*$ is induced~\cite{Lahijanian:TAC:2015}.
Our objective can then be translated as finding an optimal $\str^*_\I$ that is robust to all uncertainties induced by abstraction, i.e., 
\begin{align}
    \str_{\I}^* = \arg \max_{\str_\I} \min_{\nu_\I} Pr(\omega_\I \models \varphi \mid \str_\I, \nu_\I, \omega_\I(0) = q)
\end{align}
\noindent
$\str_{\I}^*$ can then be computed using off-the-shelf tools with a time complexity polynomial in $Q$~\cite{Lahijanian:TAC:2015}.

We can then define $\str$ according to $\str^*_\I$ by using a mapping between trajectories of System~\eqref{true_dynamics} and paths of $\I$.  Let $\map: X \rightarrow \bar{Q}$ be a mapping such that $\map(x) = q$ for all $x \in q$. With an abuse of notation, for a finite trajectory $\omega_\px \in X^*$ with length $N$, we define $\map(\omega_\px) = \map(\omega_\px(0)) \ldots \map(\omega_\px(N)) \in Q^*$.
Then, the control strategy of System~\eqref{true_dynamics} is given by: 
\begin{equation}
    \label{eq:optimal str}
    \str (\omega_{\px}) = \str_\I^*(\map(\omega_\px)).
\end{equation}
Furthermore, for $\str_\I^*$, we also obtain lower and upper bound probabilities of satisfaction of $\varphi$ from every $q \in \bar{Q}$ as 
\begin{align*}
    &\check{p}(q) = \min_{\nu_\I} Pr(\omega_\I \models \varphi \mid \str_\I^*, \nu_\I, \omega_\I(0) = q), \\
    &\hat{p}(q) = \max_{\nu_\I} Pr(\omega_\I \models \varphi \mid \str_\I^*, \nu_I, \omega_\I(0) = q).
\end{align*}
In the following theorem, we show that these probability bounds also hold for System~\eqref{true_dynamics}.

\begin{theorem}[Correctness]
\label{thm:Correctness}
    For $q \in Q$, let $\check{p}(q)$ and $\hat{p}(q)$ the lower- and upper-bound probabilities of satisfying $\varphi$ from $q$. Then, it holds that $$P^{x_0}(\omega_\px \models \varphi \mid D, \str, x_0 \in q) \in [\check{p}(q), \hat{p}(q)].$$
\end{theorem}
\begin{proof}
   $\jth{\px}(k + 1) = \jth{f}(x, a) + \jth{\pv}(k)$ is a  Gaussian process with zero mean and covariance $k_{dkl}(x,x) + \V^{(j,j)}$. Consequently, for  $x_1, ..., x_l \in D_a$ the joint distribution of $f(x_1, a) + \pv(k),...,f(x_1, a) + \pv(k)$ is still Gaussian. Then the transition kernel $T_a(q \mid x)$ in~\eqref{Eqn:Kernel} defines the one step dynamics of System~\eqref{true_dynamics}. Then, for any strategy $\str$, the upper and lower bound probabilities returned by the IMDP from initial region $q$ as built in Sec.~\ref{sec:method} contains $P^{x_0}(\omega_\px \models \varphi \mid D, \str, x_0 \in q)$ as follows from \cite[Theorem 2]{jackson2021formal}.
\end{proof}

\paragraph*{Refinement}
Recall that abstraction $\I$ relies on a discretization of $X$.  The uncertainty induced by the discretization may result in undesirable results where large sections of the space have large gap between $\check{p}$ and $\hat{p}$.  We consider a refinement strategy similar to that in \cite{Dutreix2018,Adams:CSL:2022} to efficiently reduce this conservatism. In particular, to decide on which states to refine, we define a scoring function $\beta : \bar{Q} \rightarrow \reals_{\geq 0}$ as
\begin{align*}
    \beta(q) = (\hat{p}(q) - \check{p}(q)) \sum_{a \in U}\sum_{q' \in q} (\hat{P}(q, a, q') - \check{P}(q, a, q')).
\end{align*}
$\beta$ gives higher score to states that have the most uncertainty associated with satisfying $\varphi$ and states with conservative outgoing transition probabilities. 
We refine the $n_\text{ref}$ states with the highest score and for each state we only split in half the dimension that minimizes the volume of $Z_{q,a}$ (i.e. conservatism induced by the NN linear relaxation).

\section{Case Studies}
    \label{sec:experiments}

We evaluate our DKL control synthesis framework on various nonlinear systems and cases studies.  First, we assess the learning performance of DKL under different NN architectures against other GP-based methods. 
Then, we show the efficacy of our control synthesis framework in various environments and specifications.

All experiments were run on an Intel Core i7-12700K CPU at 3.60GHz with 32 GB of RAM limited to 10 threads.  Our tool is available on GitHub\footnote{\url{https://github.com/aria-systems-group/Formal-Deep-Kernel-Synthesis}}.

\subsection{Setup and Training}
We consider three nonlinear systems from~\cite{Jackson2021, Adams:CSL:2022} as shown in Table~\ref{tab:setup}.
To learn their dynamics, we used seven learning models:
\begin{itemize}
    \item \gp: the squared exponential kernel trained on $\Dpred$.
    \item \nngp: joint NN and GP model, where the NN is trained as a predictor of the dynamics on $D_a$ and a GP is regressed to predict the error of the NN from truth on $\Dpred$.
    \item \nngpl (Limited \nngp): {\nngp} except that the NN is trained only on $\Dpred$.
    \item \fd: DKL with a NN that is trained on $D_a$ and the full output of the NN is provided as an input to the base kernel (see Figure~\ref{fig:DKM_structure}), i.e, $\DKL$. The base kernel parameters are trained on $\Dpred$.
    \item \fdl (Limited \fd): similar to {\fd} except that the NN is trained only on $\Dpred$.
    \item \sd: similar to {\fd} but only the corresponding output dimension of the NN is provided as an input to the base kernel (see Figure~\ref{fig:DKM_structure}), i.e., $\SDDKL$.
    \item \sdl (Limited \sd): similar to {\sd} except that the NN is trained only on $\Dpred$.
\end{itemize}
All NNs use the ReLU activation function. 
We trained the NN portion of the DKL models with stochastic mini-batching as a scaled predictor of the dynamics and fixed the parameters before learning the kernel parameters via maximum log likelihood.
Details on the architectures and training datasets are in Table~\ref{tab:setup}.
The primary difference between {\fd} and {\sd} models is the relation between the output of the NN and the input of the kernel as illustrated in Figure~\ref{fig:DKM_structure}.

\begin{table}[t]
    \centering
     \caption{
    Overview of the considered systems with dimensionality (Dim), number of actions in $U$, number of data points collected per action $D_a$, number of data points used for posterior predictions $\Dpred \subset D_a$, and number of layers (\# L) and neurons per layer (\# N/L) of the NNs considered.}
    \scalebox{0.96}{
    \begin{tabular}{ l  c  c  r  c  c  c }
        \toprule
        System Type & Dim.  & $|U|$ & $|D_a|$ & $|\Dpred|$ & \#L & \#N/L \\ 
        \hline
        Non-linear~\cite{Jackson2021, Adams:CSL:2022} & 2D & 4 & 1,000 & 100 & 2 & 64 \\
        Dubin's Car~\cite{Adams:CSL:2022} & 3D & 7 & 10,000 & 400 & 2 & 128\\
        2nd-order Car~\cite{Adams:CSL:2022} & 5D & 3 & 50,000 & 250 & 3 & 64 \\
        \bottomrule
    \end{tabular}
    }
    \label{tab:setup}
\end{table}

\subsection{Accuracy of Deep Kernel Learning}
We first demonstrate the advantages of DKL by comparing the predictive accuracy of the learning models above. 
We define the predictive mean and variance error of each model at a point $x$ under action $a$ as 
\begin{align*}
    &\err_\mu(x,a) = \|\expect(f(x, a)\mid D) - f(x, a)\|_{2},\\
    &\err_\sigma(x,a) = \trace(cov(f(x, a)\mid D))^\frac{1}{2},
\end{align*}
respectively.  
Table \ref{table:predictive_results} shows the maximum error values over 100,000 test points for a fixed action. 

\begin{figure}[t]
    \centering
    \begin{subfigure}[b]{0.49\linewidth}
        \includegraphics[width=\linewidth]{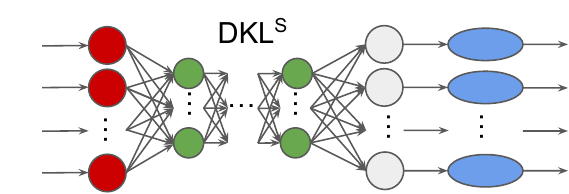}
    \end{subfigure}
    \begin{subfigure}[b]{0.49\linewidth}
        \includegraphics[width=\linewidth]{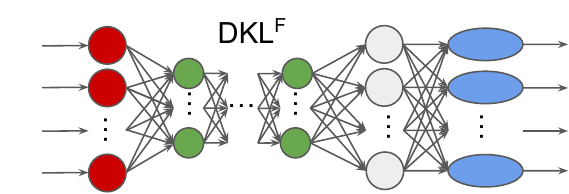}
    \end{subfigure}
    \caption{Illustration of deep kernel models. Red: NN input layer, Green: hidden layers, Grey: NN output layer, Blue: base kernel. Left: {\sd} provides single dimensional input to the kernel $k$. Right: {\fd} provides all outputs of the NN to each kernel.}
    \label{fig:DKM_structure}
\end{figure}

In all cases, {\gp} has the worse performance in mean error and compensates with large variance.  For low dimensional systems, {\nngp} performs well in mean error ($\err_\mu$) but retains a large uncertainty ($\err_\sigma$) due to poor correlation between data points in the GP.  Also, as number of dimensions increases, its predictions become drastically more uncertain.  
The DKL models have lower uncertainty ($\err_\sigma$) than {\gp} and {\nngp} models across the board, and among DKL models, {\sd} generally has the best performance (small $\err_\sigma$).  As the number of dimensions increases, the advantages of the DKL method become more significant in both mean error and variance. This is mainly due to the NN used in the prior for the kernel, which improves both mean and variance accuracy, whereas the NN in {\nngp} only improves the predicted mean, not the variance as the {\gp} prior contains insufficient information. Among NN architectures for {\fd} and {\sd} models, {\sd} (Figure~\ref{fig:DKM_structure} left) provides the best performance. This is because the NN captures the correlation between dimensions sufficiently well resulting in only the corresponding NN output being required for the base kernel to predict accurately. 

In terms of effect of data on training, {\fd} and {\sd}, where more data was used to train the NN, perform better than {\fdl},{\sdl}.  This shows that more data leads to a more accurate NN prior for the kernel. Nevertheless, over-fitting could also be a concern.  As noted in Sec.~\ref{sec:modeling_DKL}, an ill-formed prior may result in uncertainty being underestimated. We find that this is more likely to happen with low dimensional systems where a standard GP already performs sufficiently well. For example, the {\sd} with the proposed NN architecture underestimates uncertainty for one of the four modes in the 2D system, i.e., only 21\% of predictions contained the true dynamics within 2 standard deviations. However, by adding another layer to the NN or altering the training parameters, we can remove this artifact and maintain $>$95\% of predictions containing truth within two standard deviations.

\begin{table}[t]
    \centering
    \caption{
    Maximum predictive mean and variance errors over 100,000 test points for a fixed action of each system.
    }
    \label{table:predictive_results}
    \scalebox{0.96}{
    \begin{tabular} { l | c  c | c  c | c c}
        \toprule
        \multirow{2}{*}{Model}&  \multicolumn{2}{c}{\underline{\quad 2D System \quad}} & \multicolumn{2}{c}{\underline{\quad 3D System \quad}} & \multicolumn{2}{c}{\underline{\quad 5D System \quad}}\\
          &  $\err_\mu$ & $\err_\sigma$ & $\err_\mu$ & $\err_\sigma$ & $\err_\mu$ & $\err_\sigma$ \\
        \hline
        \gp & 0.6777 & 0.3473 & 1.8517 & 0.5261 & 0.7479 & 0.7825 \\
        \nngp & \textbf{0.0591} & 0.3215 &  0.1158 & 0.5209 &  0.2856 & 0.7797\\
        \nngpl & 0.0914 & 0.3218 &  0.1297 & 0.5214 & 0.2655 & 0.7794\\
        \fd & 0.2575 & 0.1817 &  0.2141 & 0.1643 & 0.1909 & 0.2626\\
        \fdl & 0.2769 & 0.1854 &  0.2068 & 0.2645 & 0.2654 & 0.2590\\
        \sd & 0.1716 & \textbf{0.1276} & \textbf{0.0856} & \textbf{0.1545} & \textbf{0.1294} & 0.1778\\
        \sdl & 0.1960 & 0.1373 &  0.1727 & 0.1575 & 0.2324 & \textbf{0.1775}\\
        \bottomrule
    \end{tabular}
    }
\end{table}

\begin{figure*}
    \centering
    \begin{subfigure}[b]{0.05\textwidth}
    \end{subfigure}
    \begin{subfigure}[b]{0.31\textwidth}
        \centering
        \qquad \; \sd
    \end{subfigure}
    \begin{subfigure}[b]{0.31\textwidth}
        \centering
        \qquad \;
        \fd
    \end{subfigure}
    \begin{subfigure}[b]{0.31\textwidth}
        \centering
        \qquad 
        \gp
    \end{subfigure}

    \begin{subfigure}[b]{0.05\textwidth}
        \centering
        {\small 0 Ref. \\
        \vspace{5mm}
        }
    \end{subfigure}
    \begin{subfigure}[b]{0.31\textwidth}
        \includegraphics[width=\textwidth, trim={0 4mm 0 0}]{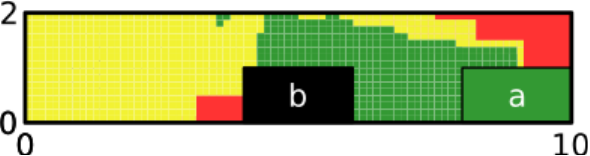}
    \end{subfigure}
    \begin{subfigure}[b]{0.31\textwidth}
        \includegraphics[width=\textwidth, trim={0 4mm 0 0}]{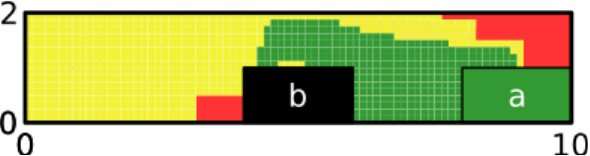}
    \end{subfigure}
    \begin{subfigure}[b]{0.31\textwidth}
        \includegraphics[width=\textwidth, trim={0 4mm 0 0}]{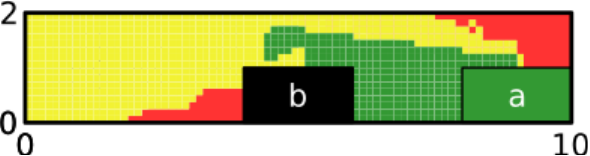}
    \end{subfigure}

    \begin{subfigure}[b]{0.05\textwidth}
        \centering
        {\small 2 Ref. \\
        \vspace{7mm}
        }
    \end{subfigure}
    \begin{subfigure}[b]{0.31\textwidth}
        \includegraphics[width=\textwidth]{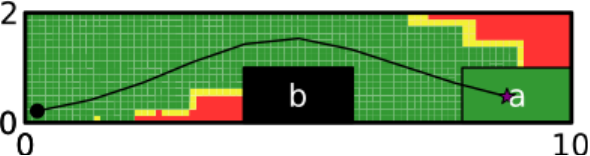}
    \end{subfigure}
    \begin{subfigure}[b]{0.31\textwidth}
        \includegraphics[width=\textwidth]{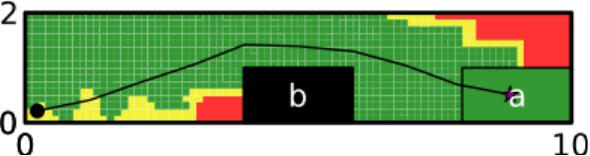}
    \end{subfigure}
    \begin{subfigure}[b]{0.31\textwidth}
        \includegraphics[width=\textwidth]{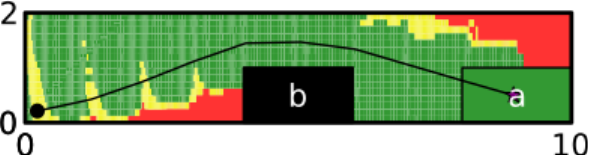}
    \end{subfigure}
    \caption{Region labeling and lower bound satisfaction probabilities $\check{p}(q)$ for experiment 1. Left: \sd, middle: \fd, and right: \gp. Top: the initial abstraction, Bottom: after two refinements. Green: $Q^{yes}$, yellow: $Q^?$, red: $Q^{no}$.}
    \label{fig:refinement}
\end{figure*}

\begin{table*}
    \centering
    \caption{Synthesis results for four different models of the 3D system. Reported values are percent volume of space for $Q^{yes/no/?}$ for the initial abstraction and after two refinements. We also report the time (minutes) taken for the NN linear relaxation, kernel bounding, transition probability calculation, and synthesis. Note that the transition probability times include the time taken to write the values to a file. Note that we recalculate the transition probabilities for every state during refinement, resulting in longer times as more states are added.}
    \label{table:3d_rs_unique}
    \begin{tabular} { l  c  c  c  c  c  c  c  c  c  r }
        \toprule
        \multirow{2}{*}{Model}& \multirow{2}{*}{\# Ref.} & \multirow{2}{*}{$|Q|$} & \multirow{2}{*}{$Q^{yes}$} & \multirow{2}{*}{$Q^{no}$} & \multirow{2}{*}{$Q^{?}$} &\multicolumn{5}{c}{\underline{\hspace{34mm} Time (min.) \hspace{34mm}}} \\
                                                    & & & & & & NN Lin. Rel. & Kernel Bounds & Trans. Prob. & Synthesis & Total \\
        \midrule
        \multirow{2}{*}{\gp} & 0 & 20,482 & 17.67 & 40.86 & 41.47 & -- & 4,866.46 & 5.87 & 3.53 & 4,875.86 \\
                             & 2 & 40,482 & 36.67 & 43.40 & 19.92 & -- &  7.00 & 17.00 & 2.24 & 26.24 \\
        \midrule
        \multirow{1}{*}{\nngp}  & 0 & 20,482 &  -- & -- & -- & 477.45 & Time Out & -- & -- & -- \\
                                
        \midrule
        \multirow{2}{*}{\fd}    & 0 & 20,482 & 18.60 & 36.27 & 45.13 & 418.8 & 63.80 & 4.98 & 1.76 & 489.34 \\
                                & 2 & 40,490 & 38.06 & 39.04 & 22.90 & -- &  1.42 & 16.72 & 1.92 & 20.06 \\
        \midrule
        \multirow{2}{*}{\sd}    & 0 & 20,482 & 20.05 & 36.46 & 43.49 & 418.8 & 8.80 & 4.73 & 1.31 & 433.64 \\
                                & 2 & 38,885 & \textbf{42.65} & \textbf{44.08} & \textbf{13.27} & -- & \textbf{0.11} & \textbf{15.37} & \textbf{1.38} & \textbf{16.86} \\
        \bottomrule
    \end{tabular}
\end{table*}

\subsection{Synthesis Results}
Here, we illustrate the efficacy of our control synthesis framework. Our metrics are the \emph{empirical validation} of guarantees, \emph{computation time}, and \emph{percent volume} of the space from which the system under the synthesized control strategy is guaranteed to satisfy the specification with probably greater than or equal to 0.95 called $Q^{yes}$, less then 0.95 called $Q^{no}$, and the remaining states called $Q^?$.



\subsubsection{Refinement and Computation Time}
We consider the 3D Dubin's car system, with the state space representing position and orientation, and synthesize a strategy for a static overtaking scenario as shown in Figure \ref{fig:refinement}. We label the stationary car as $b$ and the goal region as $a$.  The LTLf specification $\varphi_1 = \G(\lnot b) \wedge \F(a)$ then defines the task. The control synthesis results can be seen in Table~\ref{table:3d_rs_unique} and a visualization is shown in Figure~\ref{fig:refinement}. Simulated trajectories under the strategy are shown in black lines, starting from the black dot and ending at the purple star. Note that we were unable to verify the {\nngp} model due to a time out.

We see that {\sd} has the best performance, leaving only 13.27\% of the state space volume as undetermined ($Q^?$). This is expected as the {\sd} model has the highest accuracy as shown in Table~\ref{table:predictive_results}. This also follows the prediction that the lower uncertainty in DKL models would result in a less conservative abstraction. We note that {\fd} provides guarantees on a lower volume of space than {\gp}, but is capable of achieving similar results in one tenth the time and outperforms the {\gp} on the volume of $Q^{yes}$.

We note that the computational bottleneck for {\gp} abstractions comes from bounding the kernel outputs, and both {\sd} and {\fd} significantly outperform the {\gp} model in this metric. The {\sd} bounds the kernel in the shortest time due to the single dimensional input to the kernel. The {\fd} outperforms the {\gp} in time due to the NN mapping the kernel inputs into a scaled space. This allows for all of the data points used in the kernel to provide useful information about the mean and variance, as well as producing a smaller input space to calculate the bounds, unlike the GP model. In practice, we find that scaling is more effective in larger spaces. The {\nngp} model timed out during kernel bounding, taking more than 5000 minutes to bound only three of seven modes.

In each case the final abstraction consists of roughly one tenth the number of states that a uniform discretization at the finest level would produce. The tight satisfaction probabilities we see in the final abstraction for DKL highlight the efficacy of our abstraction procedure. This method is particularly effective for the DKL models. Since the NN linear relaxation holds for every $x$ inside of a discrete region $q$, refining region $q$ allows for the re-use of the linear relaxation. This enables the DKL models to take a fraction of the time as the GP model during refinements, as the linear relaxation takes significantly longer time to compute than the kernel bounds in DKL models. The linear relaxation is also the primary contributor to the conservative mean and variance bounds of the kernel, hence refinements can result in a greater change in posterior bounds for the DKL models than the GP model.

\begin{figure}[t]
    \centering
    \begin{subfigure}[b]{0.7\columnwidth}
        \includegraphics[width=\linewidth]{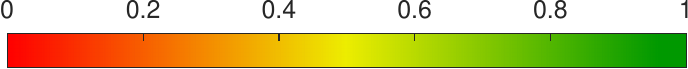}
    \end{subfigure}
    
    \begin{subfigure}[c]{0.04\columnwidth}
    \end{subfigure}
    \begin{subfigure}[c]{0.46\columnwidth}
        \centering
        \qquad \footnotesize{2D System}
    \end{subfigure}
    \begin{subfigure}[c]{0.46\columnwidth}
        \centering
        \quad \footnotesize{5D System}
    \end{subfigure}

    \begin{subfigure}[c]{0.04\columnwidth}
        \centering
        {\footnotesize Ref. 0 \\
        \vspace{5mm}
        }
    \end{subfigure}
    \begin{subfigure}[c]{0.45\columnwidth}
        \includegraphics[width=\columnwidth]{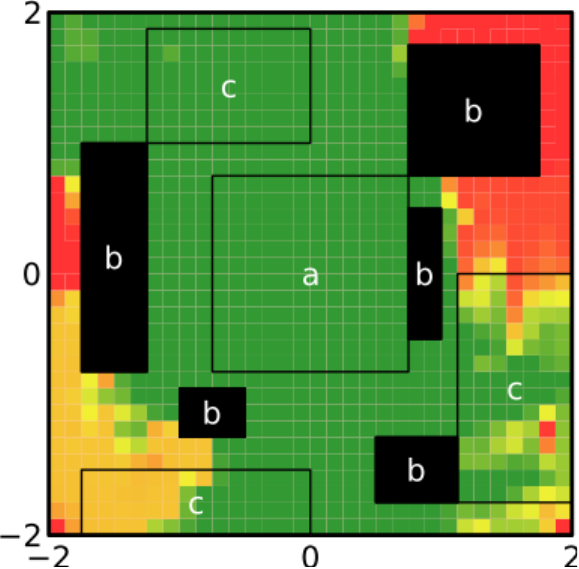}
    \end{subfigure} 
    \begin{subfigure}[c]{0.45\columnwidth}
        \includegraphics[width=\columnwidth]{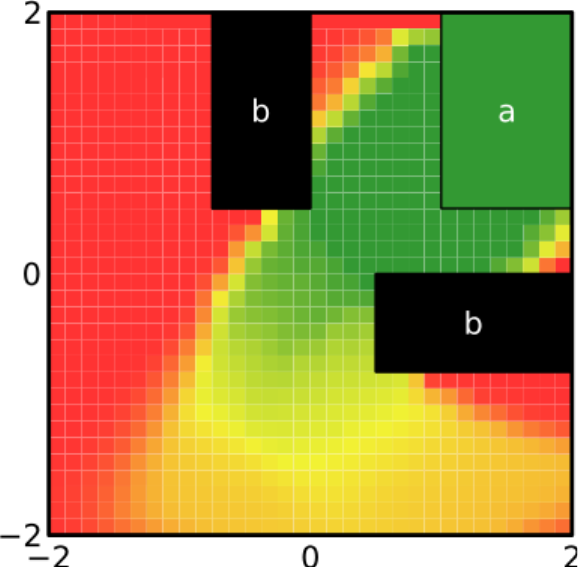}
    \end{subfigure}
    
    \begin{subfigure}[c]{0.04\columnwidth}
        \centering
        {\footnotesize Ref. 2 \\
        \vspace{5mm}
        }
    \end{subfigure}
    \begin{subfigure}[c]{0.45\columnwidth}
        \includegraphics[width=\columnwidth]{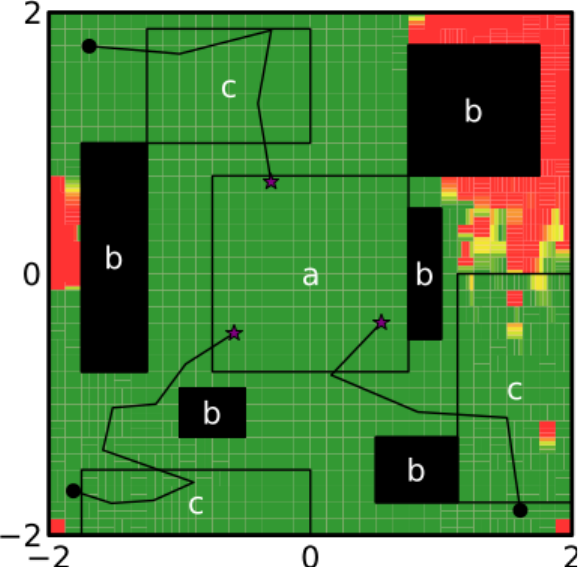}
    \end{subfigure} 
    \begin{subfigure}[c]{0.45\columnwidth}
        \includegraphics[width=\columnwidth]{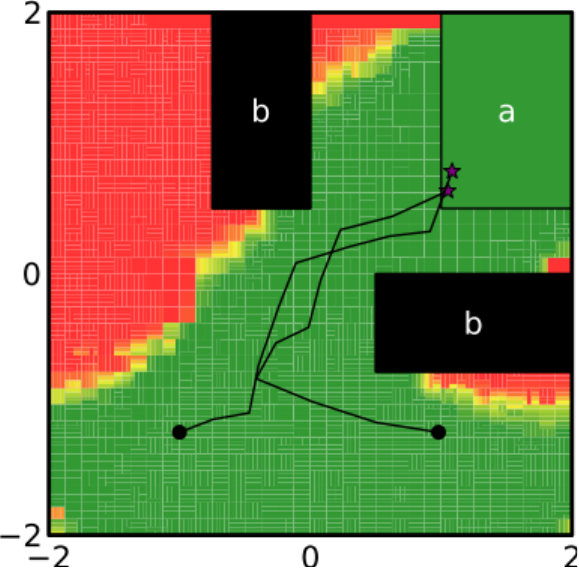}
    \end{subfigure} 

    \caption{Region labeling and lower bound satisfaction probabilities $\check{p}(q)$ for experiments 2 and 3. Left: 2D system under $\varphi_2$ and right: 5D system. Top: the initial abstraction, Bottom: after two refinements.
    }
    \label{fig:ex_1_and_3}
\end{figure}

To validate the accuracy of the satisfaction probabilities, we simulate the evolution of the system under the synthesized control strategy 1000 times from an initial region in the left half of the space, where $0.3829 \leq p(q) \leq 1.0$ and found that 100\% of simulations satisfy the specification. Similarly, simulating from a region where $0.9793 \leq p(q) \leq 1.0$, we find 100\% of simulations satisfy the specification. Simulating from an initial state where the maximum probability of satisfaction is 0 results in no trajectory remaining safe.

\subsubsection{Control Synthesis with Complex Specifications}
To show our framework can handle complex specifications, we use the 2D system and perform control synthesis given the same labeling considered in \cite{Jackson2021, Adams:CSL:2022} and LTLf specification $\varphi_2 = \G(\lnot b) \wedge \F(a) \wedge \F(c)$. In this low dimensionality, there is little difference between the synthesis results for the 4 learning models. We show results for the {\fd}. The abstraction consists of 861 states and was constructed in 10.17 minutes. We synthesized a control strategy, under which 53.22\% of the space is in $Q^{yes}$ and 26.76\% in $Q^?$, which is comparable to the results in \cite{Jackson2021, Adams:CSL:2022}. Note that \cite{Adams:CSL:2022} assumes the system model is given as a NN, whereas here we use only data and achieve similar results. After two refinements, which takes 2.4 minutes of which 1.8 are synthesis, a control strategy where 73.48\% of the space is in $Q^{yes}$ and 0.07\% in $Q^?$ is synthesized; results are shown in Figure~\ref{fig:ex_1_and_3}, simulated trajectories begin at black dots and end at purple stars. 

We validate the probabilities by simulating from the initial states of the trajectories shown in Figure~\ref{fig:ex_1_and_3} 1000 times where $0.9573 \leq p(q) \leq 1$, $0.9998 \leq p(q) \leq 1$ and $0.9916 \leq p(q) \leq 1$ find 100\% of simulations satisfy the specification, validating the bounds. 

\subsubsection{Scalability to Higher Dimensions}
We demonstrate scalability by synthesizing a control strategy for the 5D system using the environment described in \cite{Adams:CSL:2022} and LTLf specification $\varphi_1 = \G(\lnot b) \wedge \F(a)$. Here we only show results for the {\sd}, as the GP model cannot scale. The initial abstraction consists of 21,171 regions, and it took 260 minutes to calculate the NN linear relaxation and less than 6 minutes to calculate the bounds on mean and variance for the base kernel. Construction of the IMDP and synthesis took 18 minutes for the first abstraction, producing a control strategy where 17.14\% of the space is $Q^{yes}$ and 66.98\% is $Q^?$. After two refinements, which took 60 minutes with only 12 seconds being used to calculate the kernel bounds, we synthesized a control strategy where 44.36\% is $Q^{yes}$ and 39.46\% is $Q^{?}$ producing comparable results to \cite{Adams:CSL:2022} with less restrictive assumptions on the dynamics. The abstraction for this refinement consists of 42,633 states, again having one tenth the number of states of a uniform discretization; results are shown in Figure~\ref{fig:ex_1_and_3}. 

We simulate 1000 trajectories from initial states where $0.7024 \leq p(q) \leq 1.0$ and $0 \leq p(q) \leq 0.6830$ and find 100\% and 0\% of simulations satisfy the specification respectively, validating the bounds. 

\section{Conclusion}
    \label{sec:conclusion}

We introduced an abstraction framework for unknown, stochastic dynamics via DKL which can be used to synthesize strategies with guarantees on the behavior of the system. We showed the DKL models utilize the NN to transform kernel inputs into a space that enables more accurate predictions, easier computation of posterior bounds, and faster synthesis times. DKL, and our framework, enables the use of data-driven verification of systems with high-dimensionality. We note that DKL models can utilize significantly larger data sets than GPs by optimizing the NN prior over all the data and using a subset of the data for posterior predictions. This is particularly promising for systems with millions of data points available for evaluation, as this allows for a computationally tractable form of uncertainty quantification. Our abstraction procedure relies on the system having discrete modes, but recent works have provided methods for IMDP synthesis over continuous action spaces \cite{Delimpaltadakis2023}. We hope to expand our method to this domain in future work.

\bibliographystyle{IEEEtran}
\bibliography{references}

\begin{thebibliography}{10}
\providecommand{\url}[1]{#1}
\csname url@rmstyle\endcsname
\providecommand{\newblock}{\relax}
\providecommand{\bibinfo}[2]{#2}
\providecommand\BIBentrySTDinterwordspacing{\spaceskip=0pt\relax}
\providecommand\BIBentryALTinterwordstretchfactor{4}
\providecommand\BIBentryALTinterwordspacing{\spaceskip=\fontdimen2\font plus
\BIBentryALTinterwordstretchfactor\fontdimen3\font minus
  \fontdimen4\font\relax}
\providecommand\BIBforeignlanguage[2]{{%
\expandafter\ifx\csname l@#1\endcsname\relax
\typeout{** WARNING: IEEEtran.bst: No hyphenation pattern has been}%
\typeout{** loaded for the language `#1'. Using the pattern for}%
\typeout{** the default language instead.}%
\else
\language=\csname l@#1\endcsname
\fi
#2}}

\bibitem{Dutta2018}
S.~Dutta, S.~Jha, S.~Sankaranarayanan, and A.~Tiwari, ``Learning and
  verification of feedback control systems using feedforward neural networks,''
  \emph{IFAC-PapersOnLine}, vol.~51, no.~16, pp. 151--156, 2018, 6th IFAC
  Conference on Analysis and Design of Hybrid Systems ADHS 2018.

\bibitem{Haesaert2017}
S.~Haesaert, P.~M. {Van den Hof}, and A.~Abate, ``Data-driven and model-based
  verification via bayesian identification and reachability analysis,''
  \emph{Automatica}, vol.~79, pp. 115--126, 2017.

\bibitem{Jackson2021}
J.~Jackson, L.~Laurenti, E.~Frew, and M.~Lahijanian, ``Strategy synthesis for
  partially-known switched stochastic systems,'' in \emph{Proceedings of the
  24th International Conference on Hybrid Systems: Computation and Control},
  ser. HSCC '21.\hskip 1em plus 0.5em minus 0.4em\relax New York, NY, USA:
  Association for Computing Machinery, 2021.

\bibitem{Nejati2023}
A.~Nejati and M.~Zamani, ``Data-driven synthesis of safety controllers via
  multiple control barrier certificates,'' \emph{IEEE Control Systems Letters},
  vol.~7, pp. 2497--2502, 2023.

\bibitem{abate2015}
S.~Haesaert, A.~Abate, and P.~Van~den Hof, ``Data-driven and model-based
  verification: A bayesian identification approach,'' in \emph{2015 54th IEEE
  Conference on Decision and Control (CDC)}, 2015, pp. 6830--6835.

\bibitem{Knuth2021}
C.~Knuth, G.~Chou, N.~Ozay, and D.~Berenson, ``Planning with learned dynamics:
  Probabilistic guarantees on safety and reachability via lipschitz
  constants,'' \emph{IEEE Robotics and Automation Letters}, vol.~6, no.~3, pp.
  5129--5136, 2021.

\bibitem{Lederer2019}
A.~Lederer, J.~Umlauft, and S.~Hirche, ``Uniform error bounds for gaussian
  process regression with application to safe control,'' \emph{Advances in
  Neural Information Processing Systems}, vol.~32, 2019.

\bibitem{jackson2021formal}
J.~Jackson, L.~Laurenti, E.~Frew, and M.~Lahijanian, ``Formal verification of
  unknown dynamical systems via gaussian process regression,'' \emph{arXiv
  preprint arXiv:2201.00655}, 2021.

\bibitem{wajid2022formal}
R.~Wajid, A.~U. Awan, and M.~Zamani, ``Formal synthesis of safety controllers
  for unknown stochastic control systems using gaussian process learning,'' in
  \emph{Learning for Dynamics and Control Conference}.\hskip 1em plus 0.5em
  minus 0.4em\relax PMLR, 2022, pp. 624--636.

\bibitem{Tabuada2009}
P.~Tabuada, \emph{Verification and control of hybrid systems: a symbolic
  approach}.\hskip 1em plus 0.5em minus 0.4em\relax Springer Science \&
  Business Media, 2009.

\bibitem{Lahijanian:TAC:2015}
M.~Lahijanian, S.~B. Andersson, and C.~Belta, ``Formal verification and
  synthesis for discrete-time stochastic systems,'' \emph{IEEE Transactions on
  Automatic Control}, vol.~60, no.~8, pp. 2031--2045, Aug. 2015.

\bibitem{Giacomo2013}
G.~De~Giacomo and M.~Y. Vardi, ``Linear temporal logic and linear dynamic logic
  on finite traces,'' in \emph{IJCAI'13 Proceedings of the Twenty-Third
  international joint conference on Artificial Intelligence}.\hskip 1em plus
  0.5em minus 0.4em\relax Association for Computing Machinery, 2013, pp.
  854--860.

\bibitem{givan2000bounded}
R.~Givan, S.~Leach, and T.~Dean, ``Bounded-parameter markov decision
  processes,'' \emph{Artificial Intelligence}, vol. 122, no.~1, pp. 71--109,
  2000.

\bibitem{Cauchi:HSCC:2019}
N.~Cauchi, \emph{et~al.}, ``Efficiency through uncertainty: Scalable formal
  synthesis for stochastic hybrid systems,'' in \emph{Proc. of the 22nd ACM
  International Conference on Hybrid Systems: Computation and Control}, 2019,
  pp. 240--251.

\bibitem{rasmussen:book:2006}
C.~E. Rasmussen, C.~K. Williams, \emph{et~al.}, \emph{Gaussian processes for
  machine learning}.\hskip 1em plus 0.5em minus 0.4em\relax Springer, 2006,
  vol.~1.

\bibitem{liu2022}
T.~Wei and C.~Liu, ``Safe control with neural network dynamic models,'' in
  \emph{Proceedings of The 4th Annual Learning for Dynamics and Control
  Conference}, ser. Proceedings of Machine Learning Research, R.~Firoozi,
  \emph{et~al.}, Eds., vol. 168.\hskip 1em plus 0.5em minus 0.4em\relax PMLR,
  23--24 Jun 2022, pp. 739--750.

\bibitem{Adams:CSL:2022}
S.~A. Adams, M.~Lahijanian, and L.~Laurenti, ``Formal control synthesis for
  stochastic neural network dynamic models,'' \emph{IEEE Control Systems
  Letters}, 2022.

\bibitem{Mazouz:Nurips:2022}
R.~Mazouz, \emph{et~al.}, ``Safety guarantees for neural network dynamic
  systems via stochastic barrier functions,'' vol.~35, 2022, pp. 9672--9686.

\bibitem{Wilson2015}
A.~G. Wilson, Z.~Hu, R.~Salakhutdinov, and E.~P. Xing, ``Deep kernel
  learning,'' in \emph{Artificial Intelligence and Statistics}.\hskip 1em plus
  0.5em minus 0.4em\relax PMLR, 2016, pp. 370--378.

\bibitem{Ober2021}
S.~W. Ober, C.~E. Rasmussen, and M.~van~der Wilk, ``The promises and pitfalls
  of deep kernel learning,'' in \emph{Uncertainty in Artificial
  Intelligence}.\hskip 1em plus 0.5em minus 0.4em\relax PMLR, 2021, pp.
  1206--1216.

\bibitem{wang2021beta}
S.~Wang, \emph{et~al.}, ``{Beta-CROWN}: Efficient bound propagation with
  per-neuron split constraints for complete and incomplete neural network
  verification,'' \emph{Advances in Neural Information Processing Systems},
  vol.~34, 2021.

\bibitem{bertsekas1996stochastic}
D.~Bertsekas and S.~E. Shreve, \emph{Stochastic optimal control: the
  discrete-time case}.\hskip 1em plus 0.5em minus 0.4em\relax Athena
  Scientific, 1996, vol.~5.

\bibitem{pmlr-v37-sui15}
Y.~Sui, A.~Gotovos, J.~Burdick, and A.~Krause, ``Safe exploration for
  optimization with gaussian processes,'' in \emph{Proceedings of the 32nd
  International Conference on Machine Learning}, ser. Proceedings of Machine
  Learning Research, F.~Bach and D.~Blei, Eds., vol.~37.\hskip 1em plus 0.5em
  minus 0.4em\relax Lille, France: PMLR, 07--09 Jul 2015, pp. 997--1005.

\bibitem{calandra2016manifold}
R.~Calandra, J.~Peters, C.~E. Rasmussen, and M.~P. Deisenroth, ``Manifold
  gaussian processes for regression,'' in \emph{2016 International joint
  conference on neural networks (IJCNN)}.\hskip 1em plus 0.5em minus
  0.4em\relax IEEE, 2016, pp. 3338--3345.

\bibitem{deisenroth2011pilco}
M.~Deisenroth and C.~E. Rasmussen, ``Pilco: A model-based and data-efficient
  approach to policy search,'' in \emph{Proceedings of the 28th International
  Conference on machine learning (ICML-11)}, 2011, pp. 465--472.

\bibitem{Laurenti:TAC:2020}
L.~Laurenti, \emph{et~al.}, ``Formal and efficient synthesis for
  continuous-time linear stochastic hybrid processes,'' \emph{IEEE Transactions
  on Automatic Control}, 2020.

\bibitem{Patane2022}
A.~Patané, \emph{et~al.}, ``Adversarial robustness guarantees for gaussian
  processes,'' \emph{Journal of Machine Learning Research}, vol.~23, 2022.

\bibitem{Dutreix2018}
M.~Dutreix and S.~Coogan, ``Efficient verification for stochastic mixed
  monotone systems,'' in \emph{2018 ACM/IEEE 9th International Conference on
  Cyber-Physical Systems (ICCPS)}, 2018, pp. 150--161.

\bibitem{Delimpaltadakis2023}
G.~Delimpaltadakis, M.~Lahijanian, M.~Mazo~Jr, and L.~Laurenti, ``Interval
  markov decision processes with continuous action-spaces,'' in
  \emph{Proceedings of the 26th ACM International Conference on Hybrid Systems:
  Computation and Control}, 2023, pp. 1--10.

\end{thebibliography}

\end{document}